\documentclass[preprint,12pt,showpacs,preprintnumbers,amsmath,amssymb,floatfix,endfloats*]{revtex4}

\usepackage{graphicx}
\usepackage{dcolumn}
\usepackage{bm}
\usepackage{amsfonts}

\DeclareMathAlphabet{\mathsfsl}{OT1}{cmr}{bx}{it}
\begin{document}


\title{Modeling the combined effect of surface roughness and shear rate on slip flow of simple fluids}

\author{Anoosheh Niavarani and Nikolai V. Priezjev}

\affiliation{Department of Mechanical Engineering, Michigan State
University, East Lansing, Michigan 48824}

\date{\today}

\begin{abstract}

Molecular dynamics (MD) and continuum simulations are carried out to
investigate the influence of shear rate and surface roughness on
slip flow of a Newtonian fluid. For weak wall-fluid interaction
energy, the nonlinear shear-rate dependence of the intrinsic slip
length in the flow over an atomically flat surface is computed by MD
simulations. We describe laminar flow away from a curved boundary by
means of the effective slip length defined with respect to the mean
height of the surface roughness. Both the magnitude of the effective
slip length and the slope of its rate-dependence are significantly
reduced in the presence of periodic surface roughness.  We then
numerically solve the Navier-Stokes equation for the flow over the
rough surface using the rate-dependent intrinsic slip length as a
local boundary condition. Continuum simulations reproduce the
behavior of the effective slip length obtained from MD simulations
at low shear rates. The slight discrepancy between MD and continuum
results at high shear rates is explained by examination of the local
velocity profiles and the pressure distribution along the wavy
surface. We found that in the region where the curved boundary faces
the mainstream flow, the local slip is suppressed due to the
increase in pressure. The results of the comparative analysis can
potentially lead to the development of an efficient algorithm for
modeling rate-dependent slip flows over rough surfaces.

\end{abstract}


\maketitle

\section{Introduction} \label{sec:Introduction}

An adequate description of fluid flow over a solid boundary with
surface roughness on multiple length scales often requires
resolution of fine microscopic details of the flow structure while
retaining peculiarities of a macroscopic
picture~\cite{Koumoutsakos05}. The most popular and practically
important example is the problem of transport through heterogeneous
porous media, which is an inherently multiscale
system~\cite{ConvPorMed}. Modeling of the fluid flow at multiple
scales is a nontrivial problem itself; however, it becomes much more
difficult if one also needs to incorporate slip boundary conditions
into equations of motion. Typically, a local slip appears if the
wall-fluid interaction is sufficiently low, but the value of slip
length (an extrapolated distance of the velocity profile to zero)
generally depends not only on the surface energy and
structure~\cite{Churaev84,Charlaix01,Granick02,Leger06,Vinograd06}
but on shear rate as
well~\cite{Granick01,Granick02,GranLang02,Breuer03,Ulmanella08}.
Therefore, a correct form of the Navier-Stokes equation can be
obtained only if the combined effect of surface roughness and shear
rate on the slip length is taken properly into account. It is
intuitively clear that the surface roughness and shear rate have an
opposite impact on the slip length. A precise evaluation of relative
contributions of these two factors is the subject of the present
study.

The boundary conditions for the flow of monatomic fluids over
atomically flat surfaces can be described in terms of the {\it
intrinsic} slip length, which is defined as the ratio of the shear
viscosity to the friction coefficient at the liquid/solid interface.
The friction is determined by the strength of the interaction
potential and molecular-scale corrugations of the crystalline wall.
Recent MD
studies~\cite{Fischer89,KB89,Thompson90,Nature97,Barrat99,Barrat99fd,Attard04,Priezjev05,Priezjev07,PriezjevJCP}
have reported values of the intrinsic slip length up to about a
hundred molecular diameters for weak wall-fluid interaction
energies. It is well established that the formation of commensurate
structures of liquid and solid phases at the interface leads to
stick boundary conditions~\cite{Thompson90,Attard04}. The slip
length decreases with increasing bulk pressure in the flow of
nonwetting simple fluids over a smooth substrate~\cite{Barrat99}.
The MD simulations have also demonstrated that, depending on the
wall-fluid interaction energy and the ratio of wall and fluid
densities, the slip length is either relatively small (below a
couple of molecular diameters) and
shear-rate-independent~\cite{Thompson90} or larger than several
molecular diameters and increases with shear
rate~\cite{Nature97,Priezjev07}. Although several functional forms
for the rate-dependent intrinsic slip length have been suggested for
monatomic fluids~\cite{Nature97,Fang05,Priezjev07,PriezjevJCP}, as
well as for polymer
melts~\cite{Priezjev04,Priezjev08,Niavarani08,LichPRL08,Priezjev09},
the rate-dependent boundary conditions were not used for modeling
slip flows in complex geometries.

In contrast to the description of the flow over smooth boundaries
(with microscopic surface roughness) by the intrinsic slip length,
it is more appropriate to characterize the flow away from
macroscopically rough surfaces by the {\it effective} slip length,
which is usually defined with respect to the location of the mean
roughness height. Recently, we investigated the behavior of the
effective slip length in laminar flow over periodically corrugated
surfaces for simple fluids~\cite{Priezjev06} and polymer
melts~\cite{Niavarani08}. Both MD and continuum simulations have
shown that the effective slip length decreases with increasing
amplitude of the surface roughness. In the continuum analysis, the
local (rate-independent) slip length is modified by the presence of
boundary curvature~\cite{Einzel90,Einzel92} and, therefore, becomes
position-dependent along the wavy wall. A direct comparison between
MD and continuum simulations at low shear rates demonstrated that
there is an excellent agreement between the effective slip lengths
for large wavelengths (above approximately 30 molecular diameters)
and small wavenumbers $ka\lesssim0.3$~\cite{Priezjev06,Niavarani08}.
In the present paper, the analysis of the flow over a periodically
corrugated surface is extended to higher shear rates and to the
situation where the local slip length at the curved boundary is a
function of both the radius of curvature and shear rate.

In the last decade, a number of studies have focused on the
development of the hybrid continuum-atomistic methods to simulate
fluid flows near smooth solid boundaries~\cite{Thompson95,Feder00},
rough walls~\cite{NieJFM04}, superhydrophobic
surfaces~\cite{GuoWei09}, corner flows~\cite{NiePF04}, and flows
near a moving contact line~\cite{Hadji99,RenE05}. In multiscale
algorithms, the time and length scales accessible in molecular
simulations are considerably extended by employing the continuum
method in the bulk region and applying the MD technique only near
the interfaces. The two methods are coupled via constrained dynamics
in an overlap region by matching mass, momentum and energy fluxes or
velocity fields. The full MD simulations are then performed to
validate the hybrid scheme. The drawback of the hybrid methods,
however, is the cumbersome procedure for coupling of the two methods
in the overlap region. The approach pursued in the present paper is
different from that in the hybrid schemes. We first compute by MD
simulations the intrinsic slip length as a function of shear rate in
the flow over an atomically smooth surface. The Navier-Stokes
equation is then solved numerically (in the whole computational
domain) for the flow over a rough surface with the rate-dependent
intrinsic slip length used as a local boundary condition. We found
that the full MD simulations recover the continuum results at low
and intermediate shear rates and small-scale surface roughness.

In this paper, the dynamic behavior of the effective slip length is
investigated in steady shear flow over a periodically corrugated
surface using molecular dynamics and continuum simulations. Both
methods show that the effective slip length is nearly constant at
low shear rates and it gradually increases at higher shear rates.
The slight discrepancy between the effective slip lengths obtained
from MD and continuum simulations at high shear rates is analyzed by
examination of the local velocity and density profiles, as well as
pressure and temperature distributions along the wavy surface. It is
also shown by MD simulations that the singular behavior of the
intrinsic slip length (in a flow over a smooth wall) at high shear
rates is suppressed by the surface roughness.

The rest of this paper is organized as follows. The details of
molecular dynamics and continuum simulations are described in the
next section. The shear rate dependence of the intrinsic and
effective slip lengths and the results of the comparative analysis
are presented in Sec.\,\ref{sec:Results}. The results are briefly
summarized in the last section.

\section{The details of numerical methods} \label{sec:details}

\subsection{Molecular dynamics simulations} \label{sec:MD_detail}

\begin{figure}[t]
\begin{center}
\includegraphics[width=10.4cm,angle=0]{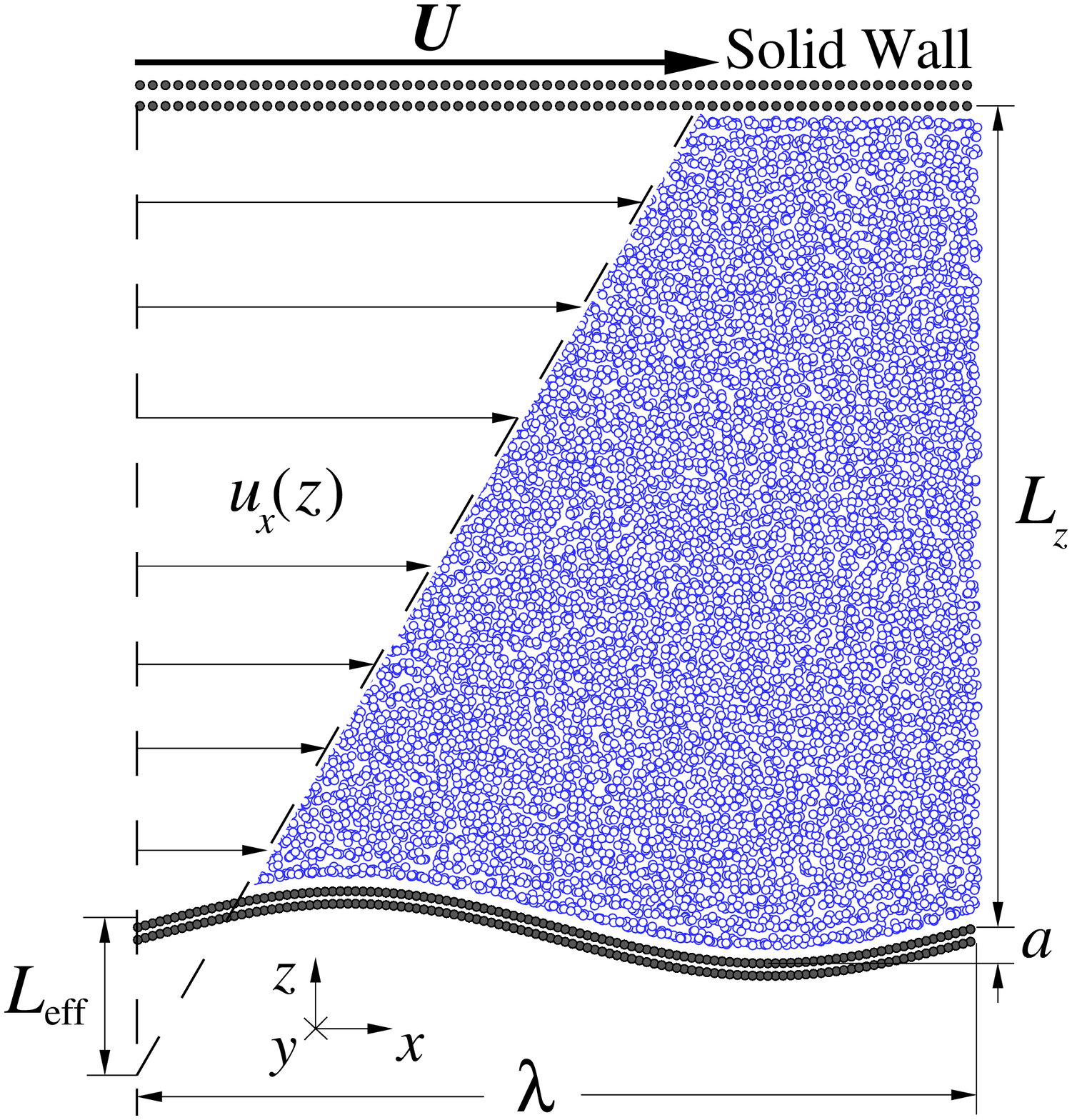}
\end{center}
\caption{(Color online) The projection of $x$ and $z$ coordinates of
the fluid monomers (open circles) confined between solid walls
(filled circles). The atoms of the lower stationary wall are
uniformly distributed along the sinusoidal curve with wavelength
$\lambda$, amplitude $a$, and density
$\rho_{w}\,{=}\,3.1\,\sigma^{-3}$. The shear flow is induced by the
flat upper wall with the density $0.67\,\sigma^{-3}$ moving with a
constant velocity $U$ in the $\hat{x}$ direction. The effective slip
length $L_{\text{eff}}$ is computed by extrapolation of the velocity
profile to $u_{x}=0$.} \label{schematic}
\end{figure}

The system geometry and the definition of the effective slip length
are illustrated in Fig.\,\ref{schematic}. The total number of fluid
monomers in the cell is fixed to $N_{f}\!=16\,104$. The fluid
monomers interact through the truncated pairwise Lennard-Jones (LJ)
potential
\begin{equation}
V_{LJ}(r)=4\varepsilon\Big{[}\Big{(}\frac{\sigma}{r}\Big{)}^{12}-\Big{(}\frac{\sigma}{r}\Big{)}^{6}\Big{]},
\label{LJ_potential}
\end{equation}
where $\varepsilon$ is the energy scale and $\sigma$ is the length
scale of the fluid phase. The cutoff radius for the LJ potential is
set to $r_{c}=2.5\,\sigma$ for both fluid-fluid and wall-fluid
interactions. The size of the wall atoms is the same as fluid
monomers. Wall atoms are fixed at the lattice sites and do not
interact with each other.

The motion of the fluid monomers was weakly coupled to a thermal
reservoir via the Langevin thermostat~\cite{Grest86}. To avoid bias
in the shear flow direction, the random force and friction terms
were added to the equations of motion in the $\hat{y}$ direction,
perpendicular to the plane of shear~\cite{Thompson90}. The equations
of motion for fluid monomers in all three directions are as follows:
\begin{eqnarray}
\label{Langevin_x}
m\ddot{x}_i & = & -\sum_{i \neq j} \frac{\partial V_{ij}}{\partial x_i}\,, \\
\label{Langevin_y}
m\ddot{y}_i + m\Gamma\dot{y}_i & = & -\sum_{i \neq j} \frac{\partial V_{ij}}{\partial y_i} + f_i\,, \\
\label{Langevin_z}
m\ddot{z}_i & = & -\sum_{i \neq j} \frac{\partial V_{ij}}{\partial z_i}\,, %
\end{eqnarray}
where $\Gamma\,{=}\,1.0\,\tau^{-1}$ is a friction coefficient that
regulates the rate of heat flux between the fluid and the heat bath,
and $f_{i}(t)$ is the random force with zero mean and variance
$2m\Gamma k_{B}T\delta(t)$ determined from the
fluctuation-dissipation relation. Unless otherwise specified, the
temperature of the Langevin thermostat is set to
$T\,{=}\,1.1\,\varepsilon/k_{B}$, where $k_{B}$ is the Boltzmann
constant. The equations of motion were integrated using the Verlet
algorithm~\cite{Allen87} with a time step $\Delta
t\,{=}\,0.002\,\tau$, where
$\tau\!=\!\sqrt{m\sigma^{2}/\varepsilon}$ is the characteristic time
of the LJ potential.


The intrinsic slip length was computed at the lower flat wall
composed out of two layers of the fcc lattice with density
$\rho_{w}\,{=}\,\,3.1\,\sigma^{-3}$ and nearest-neighbor distance
$d\,{=}\,0.77\,\sigma$ between atoms in the $xy$ plane. The partial
slip is permitted along the lower boundary due to relatively low
wall-fluid interaction energy
$\varepsilon_{wf}\,{=}\,0.5\,\varepsilon$. The no-slip boundary
condition is enforced at the flat upper wall with a lower density
$0.67\,\sigma^{-3}$ by allowing the formation of commensurate
structures of liquid and solid phases at the interface. The
nearest-neighbor distance between wall atoms is $1.28\,\sigma$ in
the ($111$) plane of the fcc lattice. The wall-fluid interaction
energy $\varepsilon$ ensures the no-slip condition at the upper wall
holds even at high shear rates. The cell dimensions are set to
$L_x\,{=}\,42.34\,\sigma$ and $L_y\,{=}\,10.0\,\sigma$ parallel to
the confining walls. The MD simulations were performed at a constant
fluid density ($\rho\,{=}\,0.81\,\sigma^{-3}$) ensemble, i.e., the
relative distance between the walls was always fixed to
$L_z\,{=}\,41.41\,\sigma$ in the $\hat{z}$ direction. Periodic
boundary conditions were applied along the $\hat{x}$ and $\hat{y}$
directions.


The effect of surface roughness on slip flow was investigated in the
cell with a lower sinusoidal wall with wavelength
$\lambda\,{=}\,L_x$ and amplitude $a\,{=}\,1.82\,\sigma$ (see
Fig.\,\ref{schematic}). Special care was taken to make the lower
corrugated wall with a uniform density
$\rho_{w}\,{=}\,\,3.1\,\sigma^{-3}$ (by including additional rows of
atoms along the $\hat{y}$ direction) so that the lattice spacing
along the sinusoidal curve remained $d\,{=}\,0.77\,\sigma$.


The fluid monomer velocities were initialized using the
Maxwell-Boltzmann probability distribution at the temperature
$T\,{=}\,1.1\,\varepsilon/k_{B}$. After an equilibration period of
about $2\times10^{4}\,\tau$, the velocity and density profiles
across the channel were averaged for about $2\times10^{5}\,\tau$
within bins of $L_{x}\times L_{y}\times\Delta z$, where $\Delta
z\,{=}\,0.01\,\sigma$. The location and dimensions of the averaging
regions for the computation of the local velocity and density
profiles will be specified in Sec.\,\ref{sec:MD_cont_flow}.

\subsection{Continuum simulations} \label{continuum_detail}

The two-dimensional, steady-state and incompressible Navier-Stokes
(NS) equation is solved numerically using the finite element method.
The equation of motion based on these assumptions is written as
follows:
\begin{equation}
\rho\textbf{u}\cdot\nabla\textbf{u}=-\nabla
p+\mu\nabla^{2}\textbf{u}, \label{N-S}
\end{equation}
where $\textbf{u}=u\,\hat{i}+v\,\hat{\!j}$ is the velocity vector,
$\rho$ is the density of the fluid, and $p$ and $\mu$ are the
pressure field and fluid viscosity, respectively.

The penalty formulation was implemented for the incompressible
solution to avoid decoupling of the velocity and pressure
terms~\cite{Pepper}. In the penalty formulation, the continuity
equation, $\nabla\cdot\textbf{u}=0$, is replaced with a perturbed
equation
\begin{equation}
\nabla\cdot\textbf{u}=-\frac{p}{\Lambda}, \label{penalty_formula}
\end{equation}
where $\Lambda$ is the penalty parameter, which enforces the
incompressibility condition. For most practical situations, where
computation is performed with double-precision $64$ bit words, a
penalty parameter ($\Lambda$) between $10^{7}$ and $10^{9}$ is
sufficient to conserve the accuracy~\cite{Pepper}. After
substitution of the pressure term in Eq.\,(\ref{penalty_formula})
into Eq.\,(\ref{N-S}), the modified momentum equation can be
rewritten as follows:
\begin{equation}
\rho\textbf{u}\cdot\nabla\textbf{u}=\Lambda\nabla(\nabla\cdot\textbf{u})+\mu\nabla^{2}\textbf{u}.
\label{penalty_final}
\end{equation}
The Galerkin method with rectangular bilinear isoparametric elements
is used to integrate the Navier-Stokes
equation~\cite{Pepper,Niavarani09}.


The boundary conditions are applied at the inlet, outlet, and upper
and lower walls. The periodic boundary conditions are implemented
along the $\hat{x}$ direction for the inlet and outlet. The boundary
condition at the upper wall is always no-slip. A partial slip
boundary condition is applied at the lower wall. In addition to the
global coordinate system ($\hat{x}, \hat{z}$), a local coordinate
system ($\vec{t}, \vec{n}$) is defined along the lower curved
boundary with unit vectors $\vec{t}$ and $\vec{n}$ representing the
tangential and normal directions, respectively. The tangential
component of the fluid velocity in the local coordinate system can
be computed as
\begin{equation}
u_{t}=L_{0}[(\vec{n}\cdot\nabla)\,u_{t}+u_{t}/R],
\label{slip_boundary}
\end{equation}
where $R$ is the local radius of curvature and $L_{0}$ is the
\emph{intrinsic} slip length at the flat liquid/solid
interface~\cite{Einzel92}. The radius of curvature $R$ is positive
and negative for concave and convex regions, respectively. The
second-order forward-differencing scheme was used to compute
accurately the normal derivative of the tangential velocity
$(\vec{n}\cdot\nabla)\,u_{t}$ at the lower boundary. 


At the beginning of the simulation procedure, the boundary
conditions at the upper and lower walls are set to no-slip. Then,
the equations of motion are solved implicitly and the tangential
velocities at the lower boundary are updated according to
Eq.\,(\ref{slip_boundary}). In the next step, the updated velocities
at the lower boundary are used as a new boundary condition and the
equations of motion are solved again. The iterative process is
repeated until a desired accuracy is achieved. The convergence rate
of the solution was controlled by applying an under-relaxation
factor of 0.001 for the boundary nodes. For all results reported in
the present paper, the continuum simulations were performed with
$150\times150$ grid resolution. However, the accuracy of the results
was also checked by performing a set of simulations using a finer
grid $180\times180$. The maximum error in the effective slip length
due to the grid resolution is $L_{\text{eff}}/L_z=0.003$. The same
error bars were reported in the previous study of laminar flow over
a periodically corrugated surface with either no-slip or
(rate-independent) slip boundary conditions~\cite{Niavarani09}.

The averaged error of the solution is defined as
\begin{equation}
\text{error}=\Big[\sum^{N_{p}}_{i=1}\frac{\mid
\textbf{u}_{i}^n-\textbf{u}_{i}^{n+1}\mid}{\mid
\textbf{u}_{i}^{n+1}\mid}\Big]/N_{p} ,
\end{equation}
where $N_{p}$ is the number of nodes in the computational domain,
$\textbf{u}_{i}^n$ is the magnitude of the velocity at node $i$ and
time step $n$, and $\textbf{u}_{i}^{n+1}$ is the magnitude of the
velocity in the next time step. The solution is converged when
$\text{error}\leqslant 10^{-9}$ and the tangential velocities along
the lower boundary satisfy $u_{t}=L_{local}\frac{\partial
u_{t}}{\partial n}$, where the local slip length is
$L_{local}=(L_{0}^{-1}-R^{-1})^{-1}$~\cite{Einzel92}.

\section{Results}
\label{sec:Results}

\subsection{MD simulations: the intrinsic and effective slip lengths}
\label{sec:MD_simulations}

\begin{figure}[t]
\begin{center}
\includegraphics[width=15cm,angle=0]{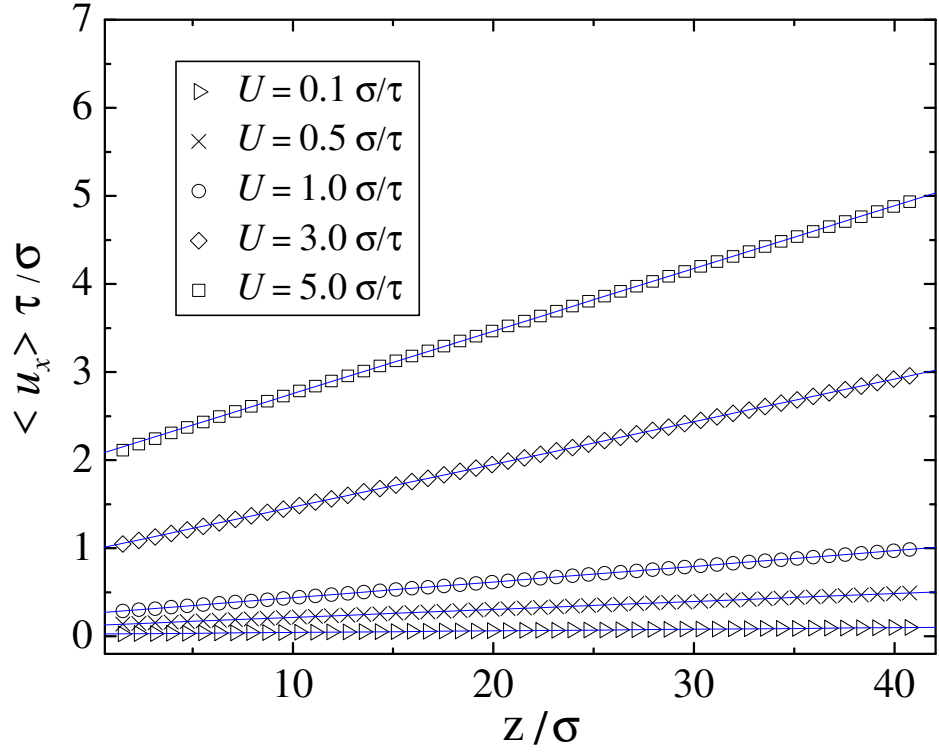}
\end{center}
\caption{(Color online) Averaged velocity profiles for the tabulated
upper wall speeds in the cell with flat upper and lower walls. The
solid lines are the best linear fits to the data. The vertical axes
indicate the location of the fcc lattice planes.} \label{vel_flat}
\end{figure}

We first consider flow in the cell with flat upper and lower walls.
The averaged velocity profiles are plotted in Fig.\,\ref{vel_flat}
for the selected upper wall speeds. The profiles are linear
throughout the channel except for a slight curvature near the lower
wall at $U\,{=}\,5.0\,\sigma/\tau$. Note that with increasing upper
wall speed, the slip velocity at the lower wall increases
monotonically up to about $2\,\sigma/\tau$ at
$U\,{=}\,5.0\,\sigma/\tau$. The flow velocity at the top boundary
remains equal to the upper wall speed due to the commensurability of
liquid and solid structures at the interface and sufficiently high
wall-fluid interaction energy. The shear rate was estimated from the
linear slope of the velocity profiles across the entire cell.

\begin{figure}[t]
\begin{center}
\includegraphics[width=15cm,angle=0]{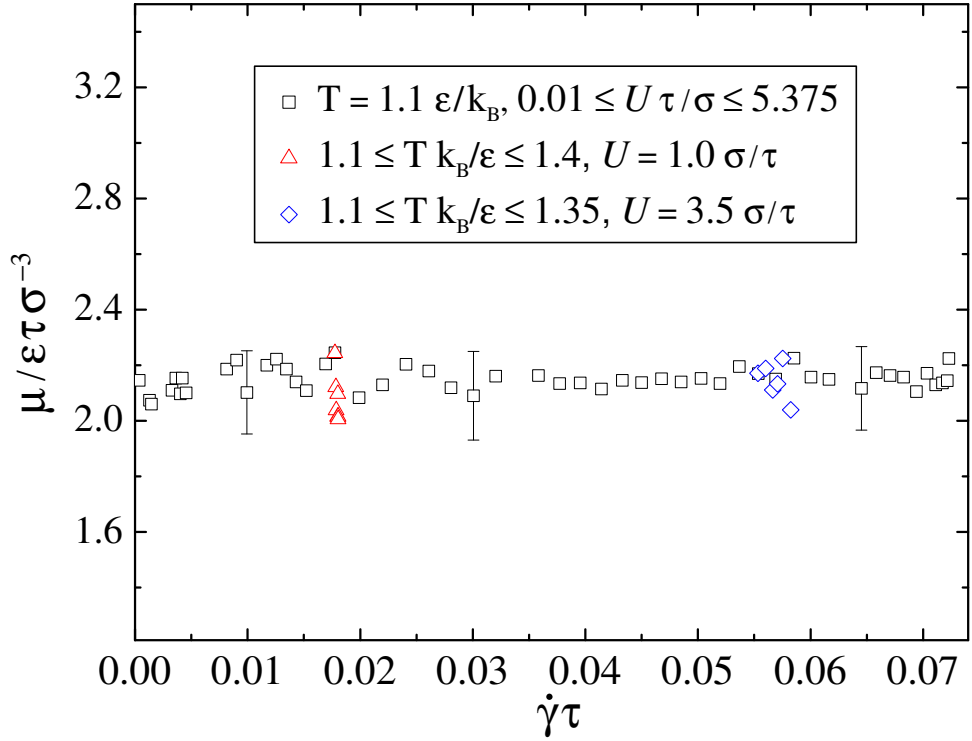}
\end{center}
\caption{(Color online) Shear viscosity in the cell with flat upper
and lower walls when the temperature of the Langevin thermostat is
set to $T\,{=}\,1.1\,\varepsilon/k_{B}$ ($\Box$). Fluid viscosity as
a function of the thermostat temperature $1.1\leqslant
T\,k_{B}/\varepsilon\leqslant1.4$ for the upper wall speed
$U\,{=}\,1.0\,\sigma/\tau$ ($\triangle$) and $1.1\leqslant
T\,k_{B}/\varepsilon\leqslant1.35$ for $U\,{=}\,3.5\,\sigma/\tau$
($\diamondsuit$).} \label{viscosity_shear}
\end{figure}

The fluid viscosity computed from the Kirkwood
relation~\cite{Kirkwood} is plotted in Fig.\,\ref{viscosity_shear}
as a function of shear rate in the cell with flat walls. In
agreement with the results of previous MD studies on shear flow of
simple fluids~\cite{Nature97,Priezjev07,PriezjevJCP}, the shear
viscosity $\mu\,{=}\,2.15\pm0.15\,\varepsilon\tau\sigma^{-3}$
remains rate-independent up to $\dot{\gamma}\tau\simeq0.072$, which
corresponds to the upper wall speed $U\,{=}\,5.375\,\sigma/\tau$. It
is also known that at high shear rates the fluid temperature
profiles become nonuniform across the cell and the heating up is
larger near the interfaces~\cite{Priezjev07,PriezjevJCP,Priezjev08}.
To ensure that the fluid viscosity is not affected by the heating
up, we performed two additional sets of simulations at the upper
wall speeds $U\,{=}\,1.0\,\sigma/\tau$ and
$U\,{=}\,3.5\,\sigma/\tau$ and increased the temperature of the
Langevin thermostat with an increment of $0.05\,\varepsilon/k_{B}$.
In the range of the thermostat temperatures reported in
Fig.\,\ref{viscosity_shear}, the shear viscosity is independent of
temperature for both upper wall speeds. In
Sec.\,\ref{sec:MD_cont_flow}, the results of simulations performed
at higher temperatures will be used to estimate the effect of
pressure on the slip length.

\begin{figure}[t]
\begin{center}
\includegraphics[width=15cm,angle=0]{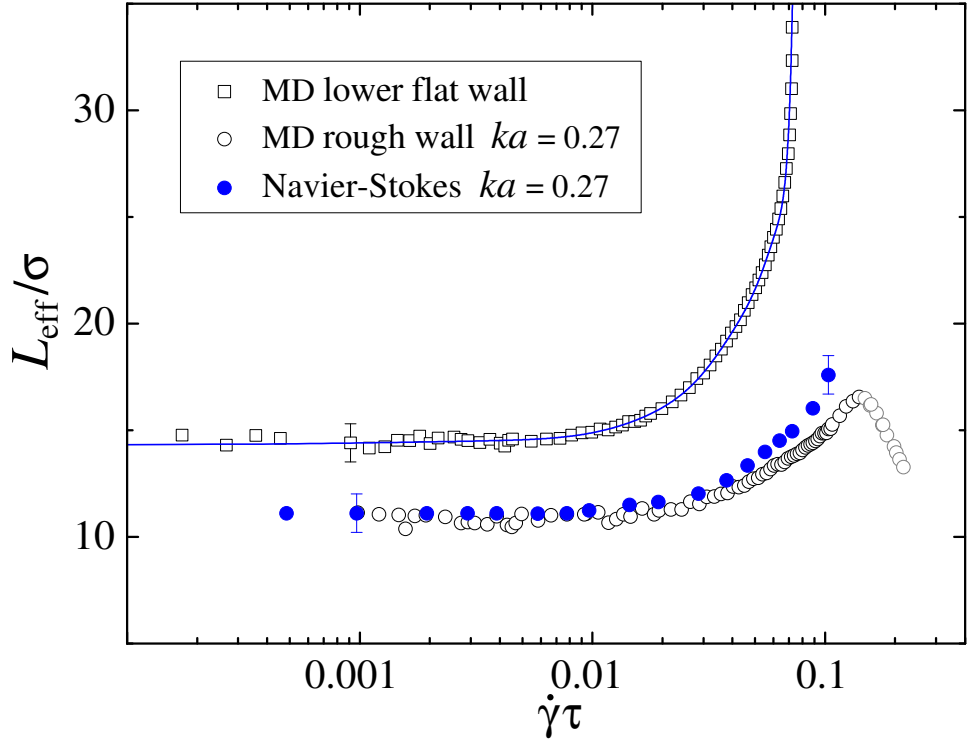}
\end{center}
\caption{(Color online) The intrinsic slip length $L_{0}/\sigma$ as
a function of shear rate for atomically flat walls ($\Box$). The
blue solid curve is a ninth-order polynomial fit to the data. The
effective slip length $L_{\text{eff}}/\sigma$ as a function of shear
rate for the flow over the corrugated wall with wavenumber $ka=0.27$
($\circ$). The effective slip length computed from the solution of
the Navier-Stokes equation with the local rate-dependent slip length
($\bullet$).} \label{sliplength_shearrate}
\end{figure}

The intrinsic slip length $L_{0}$ in the flow over a flat lower wall
was determined from the linear extrapolation of velocity profiles to
zero with respect to the reference plane located $0.5\,\sigma$ above
the fcc lattice plane. The variation of the intrinsic slip length as
a function of shear rate is presented in
Fig.\,\ref{sliplength_shearrate}. At low shear rates
$\dot{\gamma}\tau\lesssim0.005$, the slip length is nearly constant.
The leftmost data point reported in Fig.\,\ref{sliplength_shearrate}
is $\dot{\gamma}\tau\simeq1.7\times10^{-4}$ at the upper wall speed
$U\,{=}\,0.01\,\sigma/\tau$. At higher shear rates, the intrinsic
slip length first gradually increases and then grows rapidly as it
approaches $\dot{\gamma}\tau\simeq0.072$ when
$U\,{=}\,5.375\,\sigma/\tau$. The behavior of the slip length in the
flow over a flat wall can not be well described by the power law
function suggested in Ref.\,\cite{Nature97}. Instead, a ninth-order
polynomial function was used to fit the data (see the blue curve in
Fig.\,\ref{sliplength_shearrate}). In the continuum analysis
presented in the next section, the polynomial function will be used
as a local (rate-dependent) boundary condition,
Eq.\,(\ref{slip_boundary}), for the flow over a periodically
corrugated wall.


When the upper wall speed becomes greater than
$U\,{=}\,5.375\,\sigma/\tau$, the rate-dependence of the intrinsic
slip length passes through a turning point at the maximum shear rate
$\dot{\gamma}\tau\simeq0.072$, and the slip length increases
($L_0\gg L_z$) while the shear rate decreases, leading to a negative
slope of the rate-dependent curve (not shown). In this regime the
slip velocity at the lower flat wall is greater than the fluid
thermal velocity $v^{2}_{T}=k_{B}T/m$. The behavior of the intrinsic
slip length at very large slip velocities was not considered further
in the present study.

Next, the influence of surface roughness on the effective slip
length is investigated in the flow over the lower corrugated wall
with wavenumber $ka=2\pi a/\lambda=0.27$. The velocity profiles,
averaged over the period of corrugation $\lambda\,{=}\,L_x$, are
linear in the bulk of the film and curved near the lower wall within
about $2\,a$ from the bottom of the valley (e.g.,
see~\cite{Niavarani08,Niavarani09}). Therefore, the effective slip
length and shear rate were extracted from a linear fit to the
velocity profiles inside the bulk region $10\leqslant
z/\sigma\leqslant30$. The effective slip length as a function of
shear rate is also plotted in Fig.\,\ref{sliplength_shearrate}. Both
the magnitude and slope of the rate-dependence of the effective slip
length are reduced in comparison with the results for the flat wall.
Following the rate-independent regime at
$\dot{\gamma}\tau\lesssim0.01$, the effective slip length increases
monotonically up to a maximum value
$L_{\text{eff}}(0.14\,\tau^{-1})\simeq16.6\,\sigma$ at
$U\,{=}\,8.0\,\sigma/\tau$.


At higher shear rates $\dot{\gamma}\tau\gtrsim0.14$, the effective
slip length starts to decrease with further increase in shear rate
(see Fig.\,\ref{sliplength_shearrate}). We found, however, that the
fluid temperature (density) becomes anomalously high (low) on the
right side of the corrugation peak. For example, the fluid
temperature in the region $0.5\lesssim x/\lambda\lesssim0.9$ varies
up to $Tk_{B}/\varepsilon\simeq\,1.6-1.85$ at
$U\,{=}\,12.0\,\sigma/\tau$ (the rightmost point in
Fig.\,\ref{sliplength_shearrate}). Also, the amplitude of the
characteristic density oscillations normal to the curved boundary at
$x/\lambda\simeq0.43$ is reduced by about $40\%$ at
$U\,{=}\,12.0\,\sigma/\tau$ in comparison with the fluid layering at
equilibrium conditions (not shown). In the next section, the
comparison between the effective slip lengths estimated from MD and
continuum simulations is studied at shear rates
$\dot{\gamma}\tau\lesssim0.14$.




\subsection{Comparison between MD and continuum simulations}
\label{sec:comparison}

In this section, we describe the behavior of the rate-dependent
effective slip length computed from the solution of the
Navier-Stokes equation. All variables in continuum simulations are
normalized by the LJ length $\sigma$, time $\tau$, and energy
$\varepsilon$ scales. The system is designed to mimic the MD setup.
All nondimensional parameters in continuum simulations are marked by
the ($\thicksim$) sign. The cell dimensions are set to
$\tilde{L}_x\,{=}\,42.34$ and $\tilde{L}_z\,{=}\,40.41$ in the
$\hat{x}$ and $\hat{z}$ directions, respectively. Similar to the MD
setup, the wavenumber of the lower corrugated wall is fixed to
$ka=0.27$. This value was chosen based on the previous analysis of
the effective slip length in laminar flow of
simple~\cite{Priezjev06} and polymeric~\cite{Niavarani08} fluids
over a rough surface. It was shown~\cite{Priezjev06} that at low
shear rates there is an excellent agreement between the effective
slip lengths extracted from MD and continuum simulations when a
local (rate-independent) intrinsic slip length is used as a boundary
condition along the wavy wall with $\lambda\gtrsim30\,\sigma$ and
$ka\lesssim0.3$.


It is important to emphasize that in continuum simulations the
intrinsic slip length in Eq.\,(\ref{slip_boundary}) is a function of
the total shear rate at the curved boundary. The total shear rate is
computed as a sum of the normal derivative of the tangential
velocity $\partial u_{t}/\partial n$ and the ratio of the slip
velocity to the radius of curvature $u_{t}/R$. The effective slip
length is extracted from a linear part of the velocity profiles in
the bulk region $10\leqslant\tilde{z}\leqslant30$.

Figure\,\ref{sliplength_shearrate} shows the effective slip length
extracted from the continuum solution with the local rate-dependent
intrinsic slip length at the lower corrugated boundary. The upper
wall speed is varied in the range
$0.025\leqslant\tilde{U}\leqslant6$. As expected, the effective slip
length obtained from the continuum simulations agrees well with the
MD results at low shear rates $\dot{\gamma}\tau\lesssim0.01$ where
$L_0$ is nearly constant. With increasing shear rate up to
$\dot{\gamma}\tau\lesssim0.1$, the effective slip length increases
monotonically. There is an excellent agreement between the two
methods at intermediate shear rates
$0.01\lesssim\dot{\gamma}\tau\lesssim0.04$. At higher shear rates,
$0.04\lesssim\dot{\gamma}\tau\lesssim0.1$, the continuum predictions
slightly overestimate the MD results. The maximum discrepancy in the
effective slip length is about $2.5\,\sigma$ at
$\dot{\gamma}\tau\simeq0.1$, which corresponds to the upper wall
speed $\tilde{U}=6.0$. For $\tilde{U}>6.0$, the local shear rate at
the crest of the lower boundary,
$[(\vec{n}\cdot\nabla)\,u_{t}+u_{t}/R]$, becomes larger than the
highest shear rate, $\dot{\gamma}\tau\simeq0.072$, achieved in the
MD simulations for the intrinsic slip length (see
Fig.\,\ref{sliplength_shearrate}), and, therefore, the local
boundary condition, Eq.\,(\ref{slip_boundary}), cannot be applied.

\subsection{A detailed analysis of the flow near the curved boundary}
\label{sec:MD_cont_flow}

\begin{figure}[t]
\begin{center}
\includegraphics[width=15cm,angle=0]{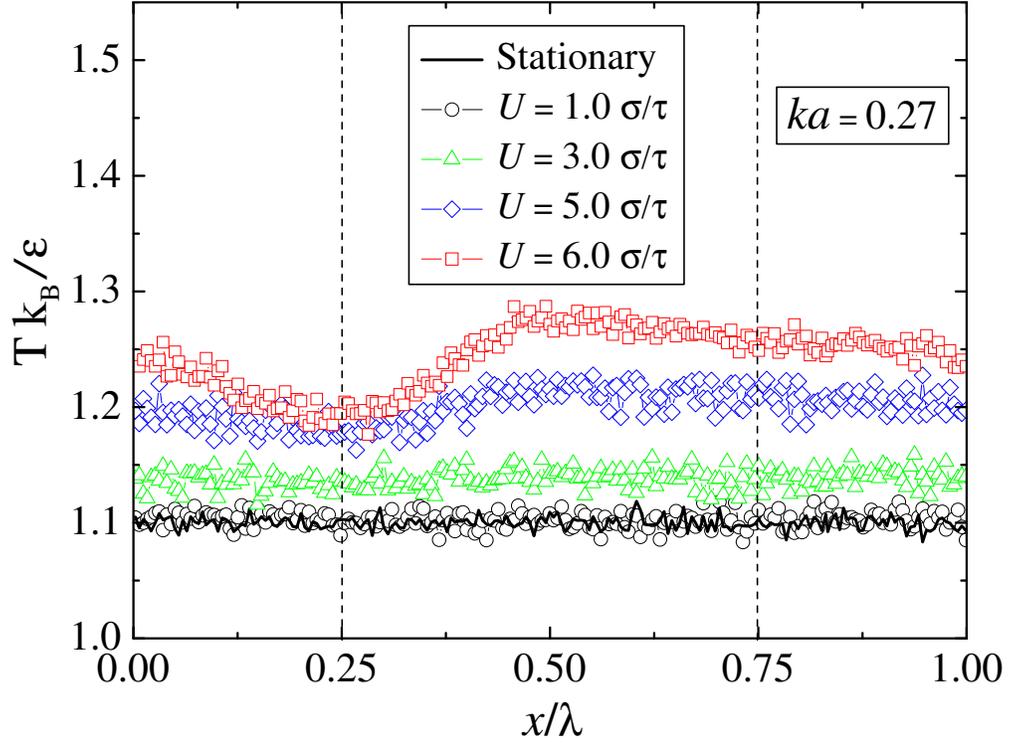}
\end{center}
\caption{(Color online) Temperature of the first fluid layer $T(x)$
(in units of $\varepsilon/k_{B}$) along the lower corrugated wall
($ka=0.27$) for the indicated upper wall speeds. The vertical dashed
lines denote the location of the crest ($x/\lambda=0.25$) and the
bottom of the valley ($x/\lambda=0.75$).} \label{temperature_rough}
\end{figure}

In order to determine what factors cause the discrepancy between the
effective slip lengths obtained from MD and continuum simulations at
high shear rates, we performed a detailed analysis of the fluid
temperature, pressure, density and velocity profiles near the lower
boundary. The fluid temperature in the first fluid layer along the
lower wall is plotted in Fig.\,\ref{temperature_rough} for the
indicated upper wall speeds. As expected, at low shear rates
$\dot{\gamma}\tau\lesssim0.019$ ($U\lesssim1.0\,\sigma/\tau$), the
fluid temperature remains equal to the value
$T\,{=}\,1.1\,\varepsilon/k_{B}$ set in the Langevin thermostat.
With increasing upper wall speed, the temperature increases and
eventually becomes nonuniformly distributed along the curved
boundary (see Fig.\,\ref{temperature_rough}). Somewhat surprisingly,
we find that at high upper wall speeds $U\geqslant5.0\,\sigma/\tau$
the fluid temperature is lowest above the crest even though the slip
velocity in this region is expected to be higher than in the other
locations.



\begin{figure}[t]
\begin{center}
\includegraphics[width=15cm,angle=0]{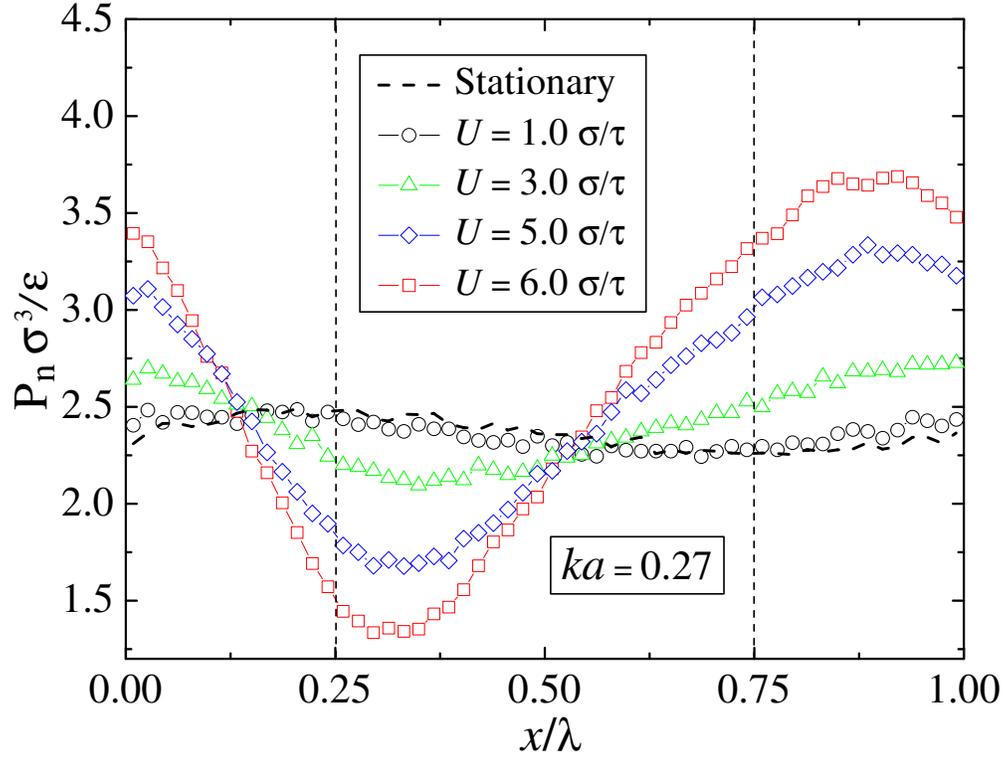}
\end{center}
\caption{(Color online) The normal pressure $P_{n}$ (in units of
$\varepsilon/\sigma^3$) along the lower corrugated wall ($ka=0.27$)
for the selected upper wall speeds. The vertical dashed lines
indicate the location of the crest ($x/\lambda=0.25$) and the bottom
of the valley ($x/\lambda=0.75$).} \label{pressure_rough}
\end{figure}

The normal pressure distribution along the lower corrugated wall is
shown in Fig.\,\ref{pressure_rough} for the selected upper wall
speeds. Each data point represents the ratio of the averaged normal
force on the wall atoms from the fluid monomers (within the cutoff
radius) to the surface area $0.77\,\sigma\times10.0\,\sigma$ in the
$ty$ plane. The small variation of the pressure at equilibrium is
due to the boundary curvature, i.e., the number of fluid monomers
within the cutoff distance from the wall atoms is on average
slightly larger above the peak than inside the valley. With
increasing upper wall speed, the normal pressure increases in the
valley ($x/\lambda\lesssim0.12$ and $x/\lambda\gtrsim0.54$) while it
becomes smaller than the equilibrium pressure on the right slope of
the peak (see Fig.\,\ref{pressure_rough}). The normal pressure
distribution extracted from the solution of the NS equation (not
shown) is qualitatively similar to the profiles presented in
Fig.\,\ref{pressure_rough}. However, in the region
$x/\lambda\gtrsim0.54$ the MD data overestimate the continuum
predictions for the pressure distribution at
$U\gtrsim3.0\,\sigma/\tau$ due to the increase in fluid temperature
(see Fig.\,\ref{temperature_rough}). It is clear from
Fig.\,\ref{pressure_rough} that at higher upper wall speeds there is
a considerable deviation from the equilibrium pressure, which can
affect the local intrinsic slip length along the curved boundary
(see discussion below). We finally comment that the normal pressure
on the flat lower wall ($ka=0$) at equilibrium is equal to
$P\simeq2.36\,\varepsilon/\sigma^{3}$ (which agrees well with the
pressure on the locally flat surface at $x/\lambda\simeq0.5$ for
$U=0$ in Fig.\,\ref{pressure_rough}) and it increases up to
$P\simeq2.52\,\varepsilon/\sigma^{3}$ at the highest upper wall
speed $U\,{=}\,5.375\,\sigma/\tau$.





\begin{figure}[t]
\begin{center}
\includegraphics[width=15cm,angle=0]{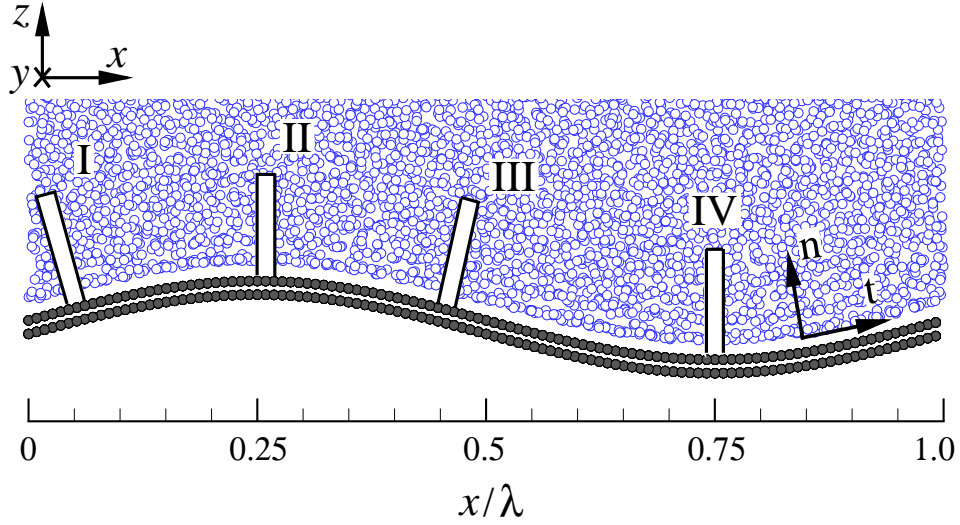}
\end{center}
\caption{(Color online) A snapshot of the fluid monomers (open
circles) near the lower corrugated wall (filled circles) with
wavenumber $ka=0.27$. The rectangular boxes denote averaging regions
on the left side of the peak (I) ($x/\lambda\simeq0.05$), above the
crest (II) ($x/\lambda\simeq0.25$), on the right side of the peak
(III) ($x/\lambda\simeq0.45$), and at the bottom of the valley (IV)
($x/\lambda\simeq0.75$). The unit vectors $\vec{t}$ and $\vec{n}$
indicate the tangential and normal directions at the boundary.}
\label{schematic_lowerwall}
\end{figure}

A snapshot of the flow near the lower corrugated wall is shown in
Fig.\,\ref{schematic_lowerwall}, where the rectangular boxes
illustrate the location of the averaging regions along the surface.
The dimensions of the boxes are $d\,{=}\,0.77\,\sigma$,
$5.0\,\sigma$, and $L_{y}\,{=}\,10.0\,\sigma$ along the curved
boundary ($xz$ plane), normal to the curved boundary ($xz$ plane),
and in the $\hat{y}$ direction, respectively. The velocity and
density profiles discussed below were averaged inside the boxes
within bins of thickness $0.01\,\sigma$ in the direction normal to
the surface.

\begin{figure}[t]
\begin{center}
\includegraphics[width=15cm,angle=0]{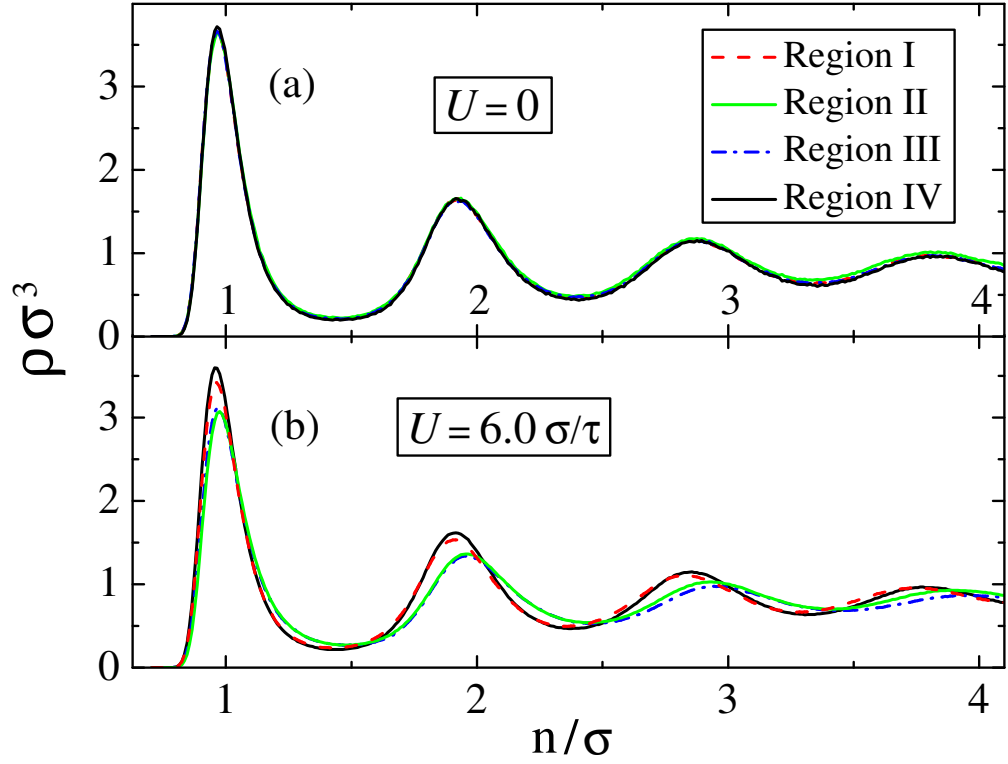}
\end{center}
\caption{(Color online) Fluid density profiles averaged inside
regions I--IV (shown in Fig.\,\ref{schematic_lowerwall}) near the
corrugated lower wall with wavenumber $ka=0.27$. The upper wall
speeds are $U=0$ (a) and $U\,{=}\,6.0\,\sigma/\tau$ (b).}
\label{density}
\end{figure}

In Figure\,\ref{density} the averaged density profiles are plotted
in regions I--IV for $U\,{=}\,0$ and $6.0\,\sigma/\tau$. In both
cases, a pronounced fluid layering is developed near the lower wall.
In the stationary case, Fig.\,\ref{density}\,(a), the density
profiles are almost the same in all four regions. Similar to the
results of previous MD studies on slip flow over smooth
surfaces~\cite{Priezjev07,PriezjevJCP}, when the upper wall speed
increases, the contact density (magnitude of the first peak) is
reduced due to a finite slip velocity along the curved boundary [see
Fig.\,\ref{density}\,(b)]. Among regions I--IV, it is expected that
the local slip velocity is highest above the crest (region II) and
lowest at the bottom of the valley (region IV), and, therefore, it
is not surprising that the contact density in the region IV is
larger than in the region II [see Fig.\,\ref{density}\,(b)].
However, this correlation does not hold in regions I and III, where
it is expected that at finite $Re$ the slip velocity in the region I
would be larger than in the region III; but, as shown in
Fig.\,\ref{density}\,(b), the contact density in the region I is
slightly larger than in the region III. Instead, we correlate this
behavior with the nonuniform pressure distribution along the curved
boundary where the local normal pressure in the region I is higher
than in the region III (see Fig.\,\ref{pressure_rough}). Note also
that the location of the second and third fluid layers in regions II
and III is slightly shifted away from the surface and the magnitude
of the peaks is reduced due to the lower pressure in the region
$0.2\lesssim x/\lambda\lesssim0.5$ at $U\,{=}\,6.0\,\sigma/\tau$ in
Fig.\,\ref{pressure_rough}.



\begin{figure}[t]
\begin{center}
\includegraphics[width=15cm,angle=0]{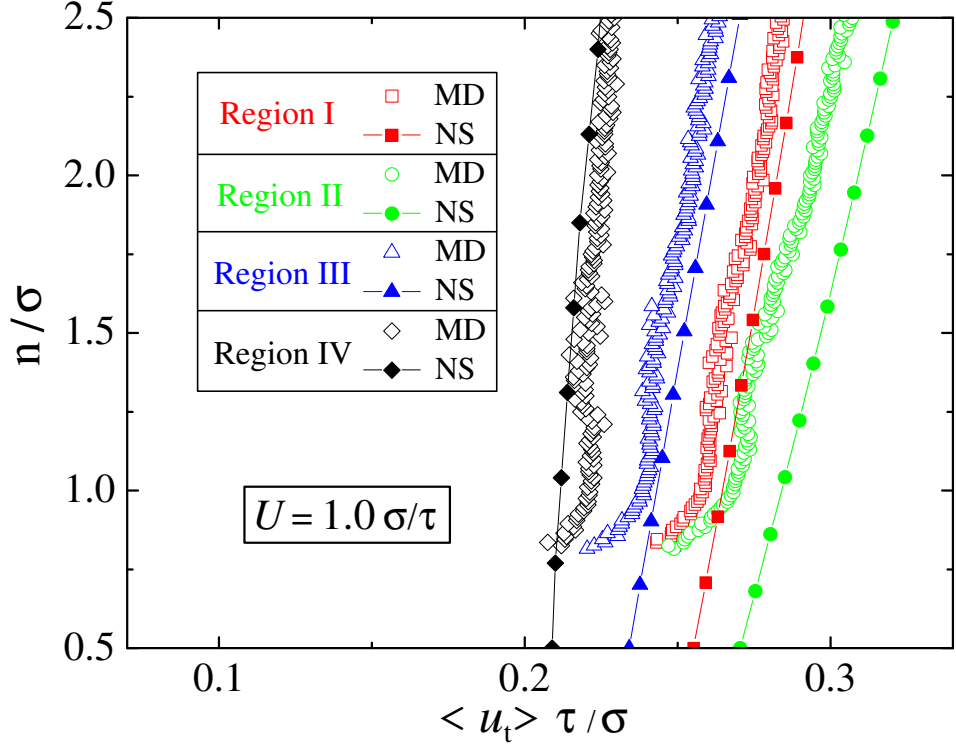}
\end{center}
\caption{(Color online) Averaged tangential velocity profiles inside
regions I--IV obtained from MD simulations (open symbols) and the
solution of the Navier-Stokes equation (filled symbols). The
location of the averaging regions near the lower corrugated wall is
shown in Fig.\,\ref{schematic_lowerwall}. The upper wall speed is
$U\,{=}\,1.0\,\sigma/\tau$.} \label{vel_comp_1.0_ut}
\end{figure}

The local tangential velocity profiles normal to the curved boundary
are presented in Fig.\,\ref{vel_comp_1.0_ut} for the upper wall
speed $U\,{=}\,1.0\,\sigma/\tau$. Note that the profiles are almost
linear away from the surface except for a slight curvature in the MD
velocity profiles near the wall. The slip velocity above the crest
(region II) is larger than in the other regions and $u_{t}$ at the
bottom of the valley (region IV) is the lowest, as expected. The
tangential velocity on the left side of the peak (region I) is
larger than on the right side (region III) due to the inertial
effects~\cite{Niavarani09}. The tangential velocity $u_{t}$ and its
normal derivative $\partial u_{t}/\partial n$ extracted from MD
simulations in regions I, III, and IV agree rather well with the
continuum results. Despite a slight overestimation of the continuum
velocity with respect to the MD results above the crest (region II),
the effective slip length is essentially the same in both methods
for the upper wall speed $U\,{=}\,1.0\,\sigma/\tau$ (see the data at
$\dot{\gamma}\tau\simeq0.019$ in Fig.\,\ref{sliplength_shearrate}).

\begin{figure}[t]
\begin{center}
\includegraphics[width=15cm,angle=0]{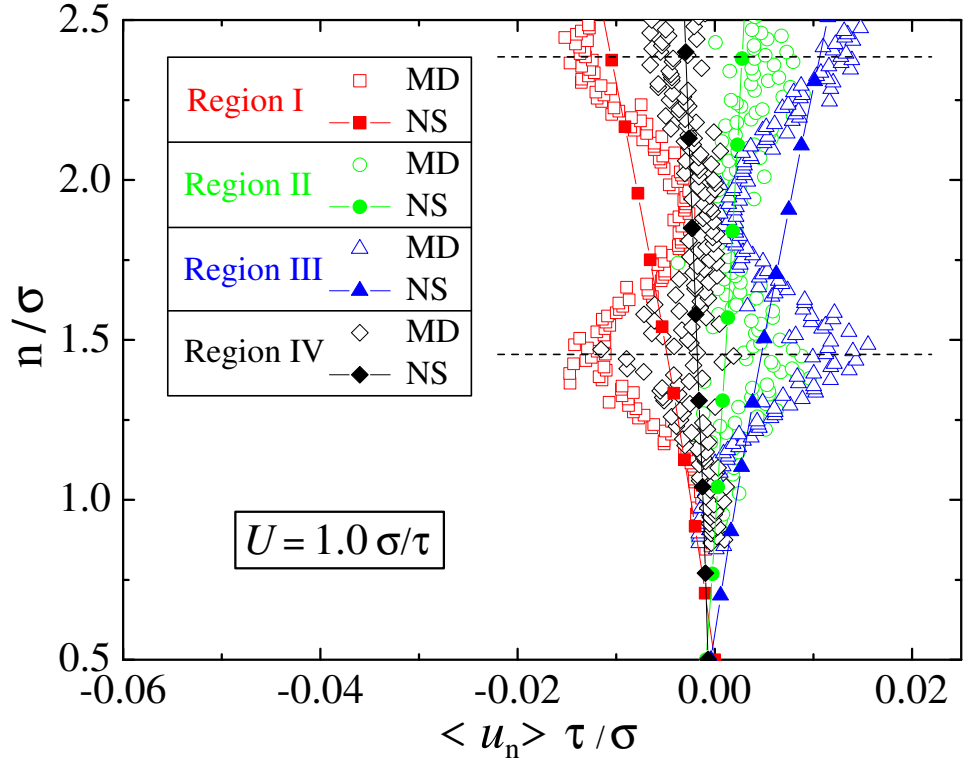}
\end{center}
\caption{(Color online) Averaged normal velocity profiles inside
regions I--IV (shown in Fig.\,\ref{schematic_lowerwall}) for the
upper wall speed $U\,{=}\,1.0\,\sigma/\tau$. The profiles are
extracted from MD simulations (open symbols) and the solution of the
Navier-Stokes equation (filled symbols). The horizontal dashed lines
indicate the location of the minima in density profiles (see text
for details).} \label{vel_comp_1.0_un}
\end{figure}

Figure\,\ref{vel_comp_1.0_un} shows local profiles of the velocity
component normal to the curved boundary for the upper wall speed
$U\,{=}\,1.0\,\sigma/\tau$. Within statistical uncertainty, the
normal velocities at the boundary are zero in both MD and continuum
simulations. In regions I and IV the normal velocity is negative,
while in regions II and III the velocity is positive away from the
boundary. The negative and positive signs of the normal velocity are
expected in regions I and III because of the positive and negative
slopes of the boundary with respect to the shear flow direction.
Although the boundary slope at the crest and at the bottom of the
valley is parallel to the upper wall, the local symmetry of the flow
with respect to $x/\lambda=0.25$ and $0.75$ is broken at finite
Reynolds numbers ($Re\approx15.6$ for $U\,{=}\,1.0\,\sigma/\tau$ in
Fig.\,\ref{vel_comp_1.0_un}), and, as a result, the normal velocity
in regions II and IV is not zero. Furthermore, the MD velocity
profiles exhibit an unexpected oscillatory pattern away from the
surface. These oscillations correlate well with the locations of the
minima among the first three fluid layers [e.g., see
Fig.\,\ref{density}(a)]. Visual inspection of the individual
molecule trajectories indicates that the fluid monomers mostly
undergo thermal motion within the moving layers but occasionally
jump back and forward between the layers. The net flux between the
layers, however, is not zero and its sign is determined by the
direction of the flow near the boundary.

\begin{figure}[t]
\begin{center}
\includegraphics[width=15cm,angle=0]{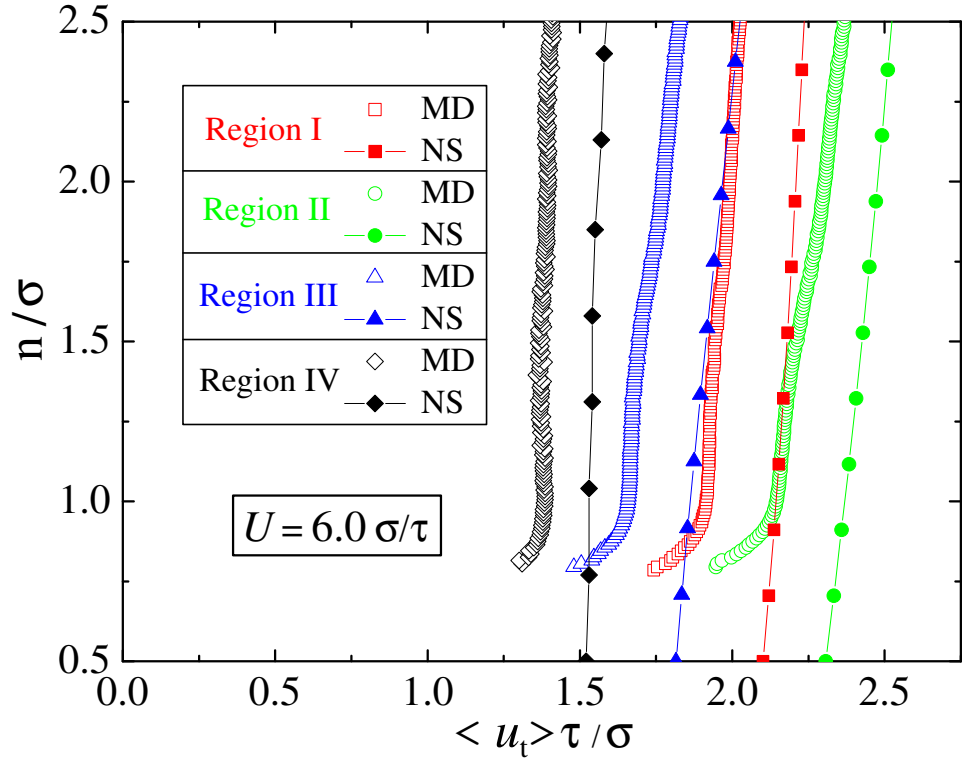}
\end{center}
\caption{(Color online) Tangential velocity profiles averaged inside
regions I--IV (shown in Fig.\,\ref{schematic_lowerwall}) near the
lower corrugated wall. The upper wall speed is
$U\,{=}\,6.0\,\sigma/\tau$. The velocity profiles are extracted from
MD simulations (open symbols) and the solution of the Navier-Stokes
equation (filled symbols).} \label{vel_comp_6.0_ut}
\end{figure}
\begin{figure}[t]
\begin{center}
\includegraphics[width=15cm,angle=0]{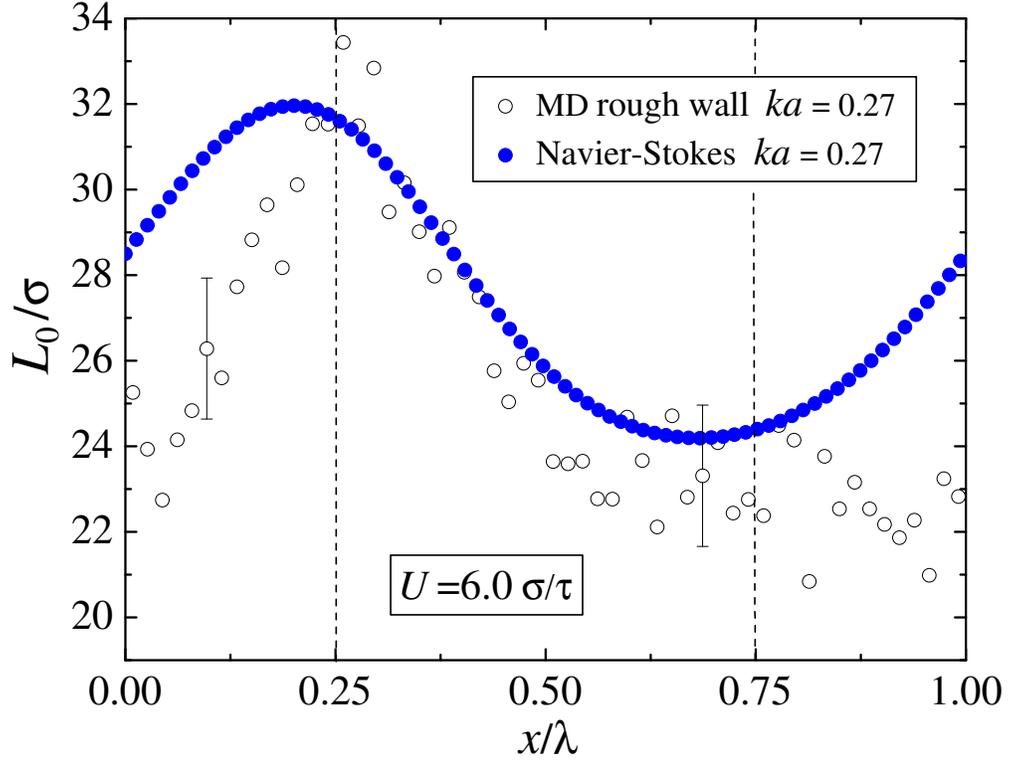}
\end{center}
\caption{(Color online) The intrinsic slip length along the lower
corrugated wall ($ka=0.27$) for the upper wall speed
$U\,{=}\,6.0\,\sigma/\tau$. The data are obtained from MD
simulations (open symbols) and the solution of the Navier-Stokes
equation (filled symbols). The vertical dashed lines denote the
position of the crest ($x/\lambda=0.25$) and the bottom of the
valley ($x/\lambda=0.75$).} \label{L0_stress_slip}
\end{figure}

The tangential velocity profiles averaged in regions I--IV at a
higher upper wall speed $U\,{=}\,6.0\,\sigma/\tau$ are shown in
Fig.\,\ref{vel_comp_6.0_ut}. The flow pattern near the lower
boundary is similar to that described in Fig.\,\ref{vel_comp_1.0_ut}
for $U\,{=}\,1.0\,\sigma/\tau$, but the relative difference between
the MD and continuum velocities is larger in each region which
results in a maximum discrepancy between the effective slip lengths
of about $2.5\,\sigma$ at $\dot{\gamma}\tau\simeq\,0.1$ in
Fig.\,\ref{sliplength_shearrate}. We tried to estimate the local
intrinsic slip length from the MD velocity profiles in regions I--IV
and to compare $L_0$ with the continuum predictions. However, the
normal derivative of the tangential velocity cannot be computed
accurately because of the curvature in the velocity profiles and
ambiguity in determining the range for the best linear fit. The
resulting error bars associated with the linear extrapolation of the
MD velocity profiles in Fig.\,\ref{vel_comp_6.0_ut} are greater than
the maximum difference between MD and continuum results for
$L_{\text{eff}}$ at high shear rates in
Fig.\,\ref{sliplength_shearrate}. Instead, a more reliable data for
the local intrinsic slip length (see Fig.\,\ref{L0_stress_slip}) are
obtained by computing the friction coefficient ($k_f=\mu/L_0$),
which is the ratio of the wall shear stress to the slip velocity.
The agreement between the MD and continuum results is quite good on
the right side of the peak in Fig.\,\ref{L0_stress_slip}. The
intrinsic slip length computed from MD simulations is about
$3-4\,\sigma$ smaller than the continuum $L_0$ on the left side of
the peak due to higher fluid pressure in that region (see also
discussion below).

\begin{figure}[t]
\begin{center}
\includegraphics[width=15cm,angle=0]{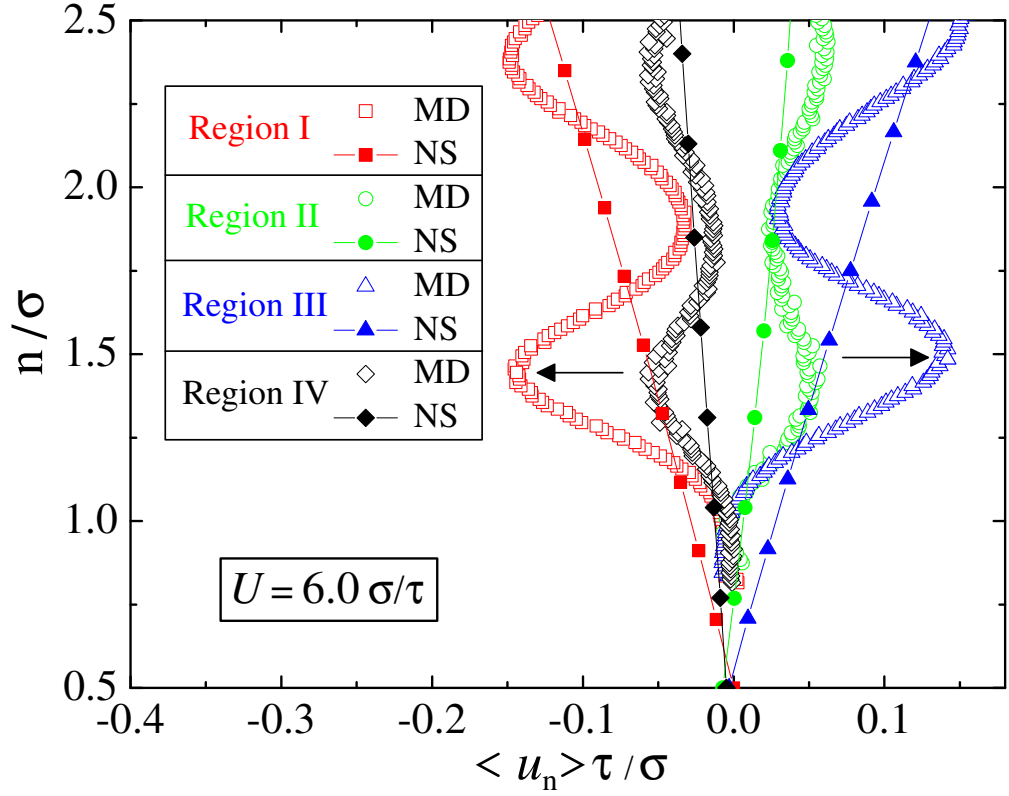}
\end{center}
\caption{(Color online) Averaged normal velocity profiles inside
regions I--IV obtained from MD simulations (open symbols) and the
solution of the Navier-Stokes equation (filled symbols). The upper
wall speed is $U\,{=}\,6.0\,\sigma/\tau$. The horizontal arrows
indicate the location of the minima in density profiles between the
first and second fluid layers in regions I and III.}
\label{vel_comp_6.0_un}
\end{figure}

The normal velocity profiles averaged in regions I--IV along the
lower boundary are displayed in Fig.\,\ref{vel_comp_6.0_un} for the
upper wall speed $U\,{=}\,6.0\,\sigma/\tau$. In each region, the
amplitude of oscillations in the MD velocity profiles is increased
with respect to the mean flow predicted by the continuum analysis.
Similar to the correlation between density and velocity profiles
observed in Fig.\,\ref{vel_comp_1.0_un} for
$U\,{=}\,1.0\,\sigma/\tau$, the locations of the maxima in the
normal velocity profiles in Fig.\,\ref{vel_comp_6.0_un} correspond
to the minima in the density profiles shown in
Fig.\,\ref{density}\,(b). At the higher upper wall speed,
$U\,{=}\,6.0\,\sigma/\tau$, however, the location of the first peak
in the velocity profile above the crest (region II) and on the right
side of the peak (region III) is slightly displaced away from the
boundary due to a small shift in the position of the second and
third fluid layers in regions II and III [see
Fig.\,\ref{density}\,(b)].

\begin{figure}[t]
\begin{center}
\includegraphics[width=15cm,angle=0]{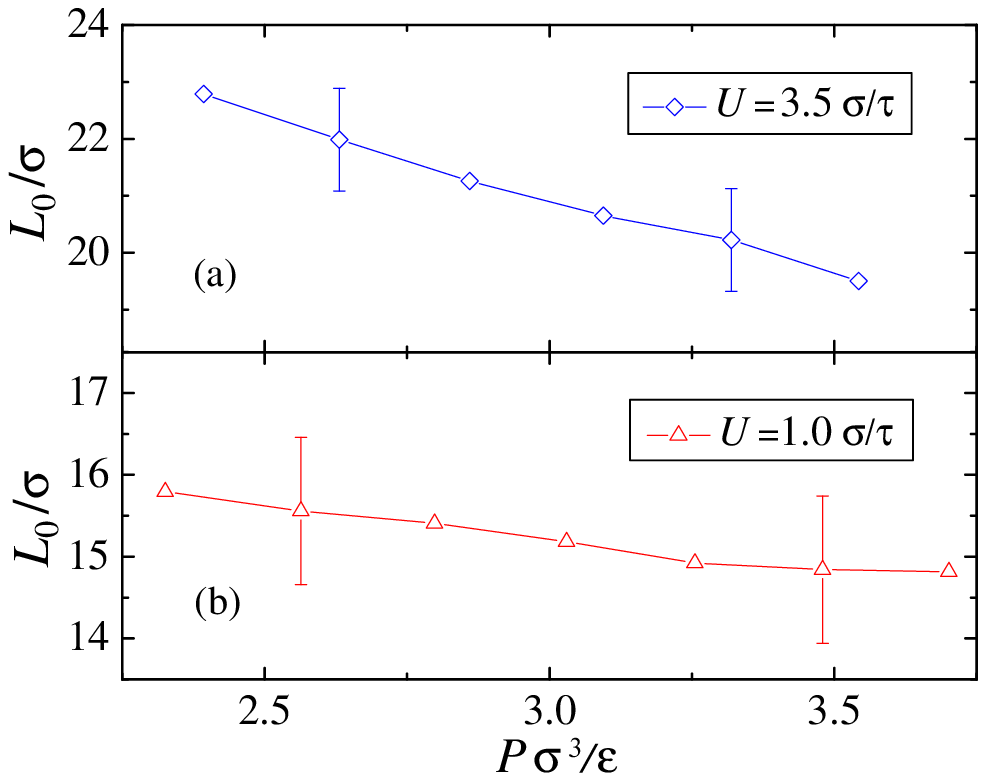}
\end{center}
\caption{(Color online) The intrinsic slip length $L_{0}/\sigma$ as
a function of the bulk pressure $P$ (in units of
$\varepsilon/\sigma^{3}$) in the cell with flat upper and lower
walls. The temperature of the Langevin thermostat is varied in the
range $1.1\leqslant T\,k_{B}/\varepsilon\leqslant1.35$ for the upper
wall speed $U\,{=}\,3.5\,\sigma/\tau$ (a) and $1.1\leqslant
T\,k_{B}/\varepsilon\leqslant1.4$ for $U\,{=}\,1.0\,\sigma/\tau$
(b).} \label{pressure_slip}
\end{figure}

The effect of pressure variation along the curved boundary is not
included in the rate-dependent intrinsic slip length,
Eq.\,(\ref{slip_boundary}), used in continuum simulations. It was
shown in Fig.\,\ref{pressure_rough} that at higher upper wall
speeds, $U\gtrsim3.0\,\sigma/\tau$, the fluid pressure along the
lower corrugated wall deviates significantly from the equilibrium
pressure. Next, we estimate the influence of pressure on the
intrinsic slip length in shear flow over a flat wall by increasing
the temperature of the Langevin thermostat. The MD simulations
described in Sec.\,\ref{sec:MD_simulations} were repeated at the
upper wall speeds $U\,{=}\,1.0\,\sigma/\tau$ and $3.5\,\sigma/\tau$
to examine $L_0$ at low ($\dot{\gamma}\tau\approx0.018$) and high
($\dot{\gamma}\tau\approx0.055$) shear rates. The results are
presented in Fig.\,\ref{pressure_slip} for the indicated
temperatures of the Langevin thermostat. As the bulk pressure
increases, the intrinsic slip length is reduced by about $3\,\sigma$
at $U\,{=}\,3.5\,\sigma/\tau$ and about $\sigma$ at
$U\,{=}\,1.0\,\sigma/\tau$. Note, however, that under these flow
conditions the shear viscosity remains constant (see
Fig.\,\ref{viscosity_shear}), and the variation of the friction
coefficient ($k_f=\mu/L_0$) is determined only by the intrinsic slip
length.

The maximum increase in pressure and the difference in $L_0$
reported in Fig.\,\ref{pressure_slip}\,(a) correlate well with the
variation of pressure in Fig.\,\ref{pressure_rough} and the
intrinsic slip length in Fig.\,\ref{L0_stress_slip} on the left side
of the peak at $U\,{=}\,6.0\,\sigma/\tau$. We conclude, therefore,
that the discrepancy between the effective slip lengths computed
from MD and continuum simulations at high shear rates (shown in
Fig.\,\ref{sliplength_shearrate}) is caused by the increase in fluid
pressure on the left side of the peak (where the solid boundary
faces the mainstream flow) and, as a result, the local intrinsic
slip length is reduced. Thus, a more accurate continuum description
of the slip flow over a curved boundary can be achieved by including
the effect of pressure and temperature on the rate-dependent
intrinsic slip length.


\section{Conclusions}

In this paper, molecular dynamics simulations were used to determine
the intrinsic slip length as a function of shear rate in steady flow
over an atomically flat surface. It was found, in agreement with
previous MD studies, that the intrinsic slip length is nearly
constant (rate-independent) at low shear rates and it grows rapidly
as the shear rate approaches a critical value. In the presence of
periodic surface roughness, the magnitude of the effective slip
length is significantly reduced and its dependence on shear rate
becomes less pronounced.

For the same surface roughness, the Navier-Stokes equation was
solved numerically with the rate-dependent boundary conditions
specified by the intrinsic slip length obtained from MD simulations.
The continuum simulations reproduce accurately the MD results for
the effective slip length at low and intermediate shear rates. The
small difference between the effective slip lengths at high shear
rates was analyzed by performing a detailed analysis of the local
velocity fields and pressure distribution along the curved boundary.
We found that the main cause of the discrepancy between MD and
continuum results at high shear rates is the reduction of the local
intrinsic slip length in the region of higher pressure where the
boundary slope becomes relatively large with respect to the
mainstream flow. These findings suggest that the continuum analysis
of the flow over a rough surface at high shear rates should take
into account the effect of pressure on the rate-dependent intrinsic
slip length.


\section{Acknowledgments}
\begin{acknowledgments}

Financial support from the Petroleum Research Fund of the American
Chemical Society is gratefully acknowledged. A. Niavarani would like
to acknowledge the generous support from the Zonta International
Foundation for the Amelia Earhart Fellowship. The molecular dynamics
simulations were conducted using the LAMMPS numerical
code~\cite{Lammps}. Computational work in support of this research
was performed at Michigan State University's High Performance
Computing Facility.

\end{acknowledgments}

\end{document}